\documentclass[aps,prd,superscriptaddress,showpacs,preprint,amsmath,amssymb]{revtex4}
\usepackage{graphicx, bm}
\usepackage{multirow}
\usepackage[usenames]{color}
\usepackage{slashed}

\usepackage{subcaption}
\usepackage{caption}
\usepackage{slashed}
\usepackage{float}
\begin{document}

\draft

\title{Analysis of anomalous $H \gamma\gamma$ coupling in light-by-light collision at future muon collider}

\author{Serdar Spor\footnote{serdar.spor@beun.edu.tr}}
\affiliation{\small Department of Medical Imaging Techniques,
Zonguldak Bülent Ecevit University, 67100, Zonguldak, Türkiye.\\}

\author{Emre Gurkanli\footnote{egurkanli@sinop.edu.tr}}
\affiliation{\small Department of Physics, Sinop University, Türkiye.\\}

\date{\today}

\begin{abstract}

In this study, the process $\gamma\gamma \to \gamma\gamma$ is investigated to establish constraints on anomalous Higgs boson couplings at $H\gamma\gamma$ vertex within the framework of the Standard Model Effective Field Theory (SMEFT). The study is performed for a future muon collider operating at CoM energies of 10 and 30 TeV with integrated luminosities of 10 and 90 ab$^{-1}$, respectively, where the incoming photons are modeled using the Weizsäcker-Williams approximation. Signal and background events are simulated using MadGraph, with the SMEFT Lagrangian implemented via FeynRules and UFO frameworks. Parton showering is evaluated with PYTHIA 8, and detector effects are accounted for using Delphes. Limits on the Wilson coefficients $\bar{c}_\gamma$ and $\tilde{c}_\gamma$ of the dim-6 operators without and with systematic uncertainties of 5\% and 10\% are reported at the 95\% confidence level, demonstrating the potential of a high-energy muon collider to provide precise constraints on these couplings and presenting a significant improvement over experimental and related phenomenological results.

\end{abstract}


\vspace{5mm}

\maketitle


\section{Introduction}

The discovery of the Higgs boson by the ATLAS \cite{Aad:2012les} and CMS \cite{Chatrchyan:2012les} experiments at the Large Hadron Collider (LHC) represents a keystone in particle physics, attesting the Higgs mechanism as the origin of the electroweak symmetry breaking (EWSB). While this discovery confirmed the completeness of the Standard Model (SM), it also brought important questions about the nature of the Higgs boson and its interactions with the existing particles.  Experimental evidence aligns with SM predictions, describing the Higgs as a CP-even scalar \cite{Aad:2016saf,Aad:2020czq,Aad:2020jkn,Sirunyan:2020wxc,Tumasyan:2022ssx}. However, the observed matter-antimatter asymmetry in the Universe suggests the existence of additional sources of CP violation which have not been discovered beyond those provided by the SM \cite{Steigman:1976ghb,Cohen:1987efv,Steigman:2008ujb}, such as the Cabibbo-Kobayashi-Maskawa (CKM) mechanism \cite{Cabibbo:1963les,Kobayashi:1973hqw}.

To address these questions, theoretical frameworks such as the Standard Model Effective Field Theory (SMEFT) offer powerful tools for systematically exploring new physics through higher-dimensional operators. Within SMEFT, CP-violating Higgs interactions with gauge bosons are of particular interest. Investigating these interactions could provide crucial insights into the origins of CP violation and the underlying mechanisms responsible for the baryon asymmetry of the Universe \cite{Riotto:1999jdx}. Precision measurements of these couplings are essential for probing deviations from SM expectations, thereby paving the way for understanding physics beyond the SM \cite{Kuzmin:1985yla}.

Future colliders, such as the muon collider, represent an opportunity to study the Higgs boson with remarkable precision. The muon collider is equipped to investigate CP-Conserving (CPC) and CP-Violating (CPV) interactions within the SMEFT framework at CoM energies of 10 and 30 TeV with integrated luminosities of 10 and 90 ab$^{-1}$, respectively. This includes processes such as $\gamma\gamma \to \gamma\gamma$, where photons are chosen under the Weizsäcker-Williams approximation \cite{Weizsacker:1934fzw,Williams:1934tsx}, which is the main focus of this study. This process, conducted by electroweak interactions at the tree level, provides an ideal platform for studying the effects of dim-6 operators on Higgs-gauge boson couplings. The high precision capability of muon collider makes it an ideal tool in the search for deviations from SM predictions.

The effective exploration of such rare processes relies on sophisticated computational and experimental methodologies. Event generation utilizing tools such as MadGraph, combined with SMEFT Lagrangian implementations via FeynRules and the UFO framework, enables precise modeling of both signal and background processes. These events are refined through parton showering with PYTHIA 8 and detector simulation with Delphes to include realistic experimental effects. By analyzing the kinematical variables of the outgoing photons, this study aims to extract robust constraints on both the CPC and CPV Wilson coefficients $\bar{c}_\gamma$ and $\tilde{c}_\gamma$, thereby offering critical insights into the Higgs boson’s interactions and their potential contributions to CP violation.

The findings are expected to provide a guidance for future experiments and theories that address open questions, such as the nature of dark matter, the hierarchy of neutrino masses, and the mechanisms behind baryon asymmetry. The sections of the study are organized as follows: Section II outlines the theoretical framework for Higgs-gauge boson couplings, while Section III details the methodology for testing these couplings at a future muon collider. The results are presented in Section IV, and concluding remarks in Section V.

\section{Effective Theory Framework for Higgs-Gauge Boson Couplings}

The SM of particle physics, structured as a quantum field theory under \(SU(3)_c \times SU(2)_L \times U(1)_Y\) gauge symmetry, describes elementary particles and their interactions. The SM Lagrangian is limited to operators with mass dimensions of four or less, maintaining Lorentz invariance and gauge symmetry. Extensions of the SM are often formulated within the framework of effective field theory, where higher-dimensional operators represent the residual effects of physics beyond the SM.

In this study, we employ the SMEFT in the strongly interacting light Higgs (SILH) framework, focusing on dimension-6 operators within the ``bar" convention as outlined in the literature \cite{Englert:2016edw, Contino:2013kra, Alloul:2014naa}. Here, the coefficients are parameterized as $\bar{c} \equiv c(M^2/\Lambda^2)$, where $M$ is a scale such as $v$ or $m_W$, and $c \sim g^2_{NP}$, with $g_{NP}$ being a new physics coupling. The effective Lagrangian for Higgs interactions that respect SM gauge symmetry is expressed as \cite{Giudice:2007ops}:

\begin{eqnarray}
\label{eq.1} 
\mathcal{L}_{\rm eff} ={\cal L}_{\text{SM}}+{\cal L}_{\text{CPC}}+{\cal L}_{\text{CPV}}= \mathcal{L}_{\rm SM} + \sum_i \bar{c}_i {\cal O}_i + \sum_i \tilde{c}_i {\cal O}_i,
\end{eqnarray}

Here, \(\bar{c}_i\) and \(\tilde{c}_i\) are the Wilson coefficients for CPC and CPV interactions, respectively, and ${\cal O}_i$ is dimension-6 operator proportional to these coefficients. This work specifically considers the interactions between the Higgs boson and electroweak gauge bosons as described in Ref.~\cite{Alloul:2014naa}. The effective Lagrangian of these interactions includes terms proportional to \(\Phi^\dagger \Phi B_{\mu\nu} B^{\mu\nu}\) for \(\bar{c}_\gamma\) and \(\Phi^\dagger \Phi B_{\mu\nu} \widetilde{B}^{\mu\nu}\) for \(\tilde{c}_\gamma\), where \(B_{\mu\nu}\) and \(\widetilde{B}_{\mu\nu}\) are the field strength tensor and its dual. In the below, we first give the CPC and CPV Lagrangian parts and then the definitions of gauge field strength tensor and its dual.

\begin{eqnarray}
\label{eq.2} 
{\cal L}_{\text{CPC}}=\frac{g^{\prime2}\bar{c}_{\gamma}}{m_W^2} \Phi^\dagger \Phi B_{\mu\nu} B^{\mu\nu}, \quad {\cal L}_{\text{CPV}}=\frac{g^{\prime 2} \tilde{c}_{\gamma}}{m_W^2}\Phi^\dagger \Phi B_{\mu\nu} \widetilde{B}^{\mu\nu}
\end{eqnarray}

\begin{eqnarray}
\label{eq.3} 
{B}_{\mu\nu}=\partial_\mu B_\nu - \partial_\nu B_\mu , \quad
\widetilde{B}_{\mu\nu}=\frac{1}{2}\epsilon_{\mu\nu\rho\sigma}B^{\rho\sigma}
\end{eqnarray}

The SILH basis for the CPC and CPV dimension-6 operators, as expressed in Eq.~(\ref{eq.2}), can be reformulated in terms of mass eigenstates following electroweak symmetry breaking. In the mass basis and under the unitarity gauge, the effective Lagrangian describing the three-point interactions involving at least one Higgs boson and two photons is structured as follows \cite{Alloul:2014naa}:

\begin{eqnarray}
\label{eq.4} 
{\cal L}=-\frac{1}{4}g_{h\gamma\gamma}F_{\mu\nu}F^{\mu\nu}h-\frac{1}{4}\widetilde{g}_{h\gamma\gamma}F_{\mu\nu}\widetilde{F}^{\mu\nu}h
\end{eqnarray} 

Here, $h$ represents the Higgs boson field. The connection between the Lagrangians in the gauge basis given in Eq.~(\ref{eq.2}) and in the mass basis given in Eq.~(\ref{eq.4}) is expressed as follows,

\begin{eqnarray}
\label{eq.5} 
g_{h\gamma\gamma}=a_H-\frac{8g\bar{c}_\gamma s_W^2}{m_W}
\end{eqnarray}

\begin{eqnarray}
\label{eq.6} 
\widetilde{g}_{h\gamma\gamma}=-\frac{8g\tilde{c}_\gamma s_W^2}{m_W}.
\end{eqnarray}

{\raggedright where $s_W = \sin\theta_W$; $\theta_W$ is the weak mixing angle. $a_H$ represents the SM contributions at the Higgs boson to two-photon vertex. Thus, the CPC and CPV dimension-6 operators in the SILH basis are described in the mass eigenstate basis after electroweak symmetry breaking and in this framework, we calculate the Wilson coefficients \(\bar{c}_\gamma\) and \(\tilde{c}_\gamma\), which describe CPC and CPV interactions of the Higgs boson with photons.}

In this study, we investigate the sensitivity of the Wilson coefficients $\bar{c}_{\gamma}$ and $\tilde{c}_{\gamma}$ in the anomalous $H\gamma\gamma$ vertex through the process $\gamma\gamma \to \gamma\gamma$ at a future muon collider. As one of the golden channels to discover the Higgs boson, the $H \to \gamma\gamma$ process has been of great interest in high energy physics due to the excellent performance of photon identification at the LHC \cite{Dong:2022hsx}, which increases the importance of the selected process in this study. The analysis focuses on evaluating the potential CPC and CPV contributions associated with these coefficients. This study employs dim-6 operators within the SMEFT framework, implemented in MadGraph5$\_$aMC@NLO \cite{Alwall:2014cvc}, with model files created using FeynRules \cite{Alloul:2014tfc} and UFO \cite{Degrande:2012acs} frameworks. In simulation setup, we have used the Weizsäcker-Williams approximation for modeling the initial photons.

Numerous phenomenological studies have been performed to constrain the Wilson coefficients of CPC and CPV dim-6 operators in various channels at \(pp\) \cite{Ellis:2015tdw,Englert:2016edw,Khanpour:2017ssa,Ferreira:2017qmj,Denizli:2019oxc,Denizli:2021uhb,Denizli:2021pkw}, \(ee\) \cite{Ellis:2014opx,Kumar:2015rvx,Ellis:2016dmk,Khanpour:2017ubw,Alam:2017rma,Ellis:2017klz,Denizli:2018rca,Karadeniz:2020yvz}, \(ep\) \cite{Kuday:2018yza,Hesari:2018ygv}, and \(\mu\mu\) \cite{Spor:2024sdk,Gurkanli:2025hju,Spor:2025eph} colliders. Muon colliders, particularly those operating at multi-TeV energy scales, provide a unique platform for probing new physics due to their cleaner environment and reduced backgrounds compared to hadron colliders. In view of their higher mass, muons produce significantly less synchrotron radiation than electrons, enabling their acceleration to higher energies in circular colliders. A muon collider at CoM energies of 10 and 30 TeV with integrated luminosities of 10 and 90 ab$^{-1}$, respectively, offers an exceptional opportunity to study anomalous Higgs interactions with gauge bosons, particularly the $H\gamma\gamma$ coupling. 

\begin{figure}[H]
\centerline{\scalebox{0.35}{\includegraphics{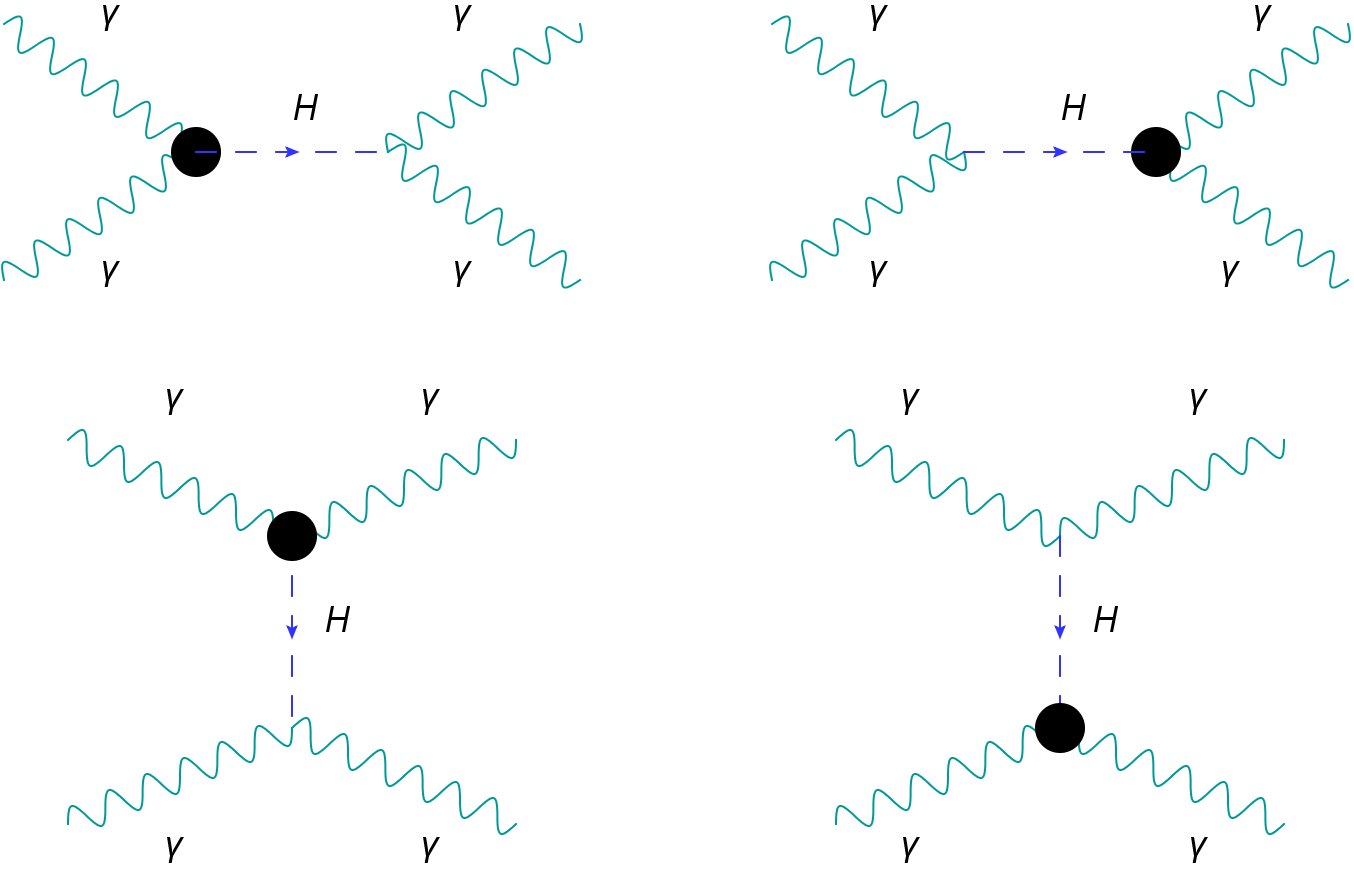}}}
\caption{ \label{fig:1}  Feynman diagrams of the $\gamma\gamma \to \gamma \gamma$ process containing the anomalous $H\gamma\gamma$ vertex.}
\end{figure}

\begin{figure}[H]
\centerline{\scalebox{0.91}{\includegraphics{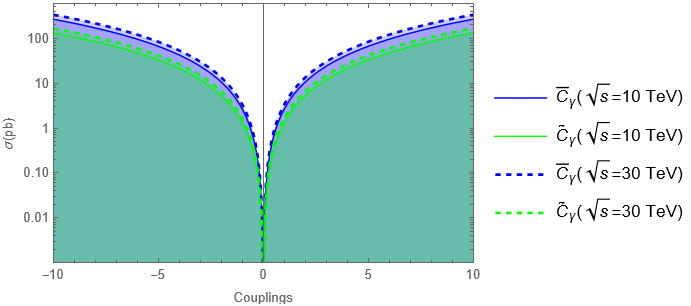}}}
\caption{ \label{fig:2} The total cross-sections of the process $\gamma\gamma \to \gamma \gamma$ in terms of the coefficients $c_{\gamma}$ and $\tilde{c}_{\gamma}$.}
\end{figure}

The Feynman diagrams of the process $\gamma \gamma \to \gamma \gamma$ that represents the new physics contributions with black dots are given in Fig.~\ref{fig:1}. In Fig.~\ref{fig:2}, the process $\gamma\gamma \to \gamma\gamma$ is examined by varying the parameters $\bar{c}_{\gamma}$ and $\tilde{c}_{\gamma}$ individually, while the other coefficients are fixed to zero, to isolate its effect. When all coefficients are set to zero, the cross-section corresponds to the SM prediction for this process. This approach allows us to calculate total cross-sections for different values of $\bar{c}_{\gamma}$ and $\tilde{c}_{\gamma}$, highlighting deviations from the SM. The results demonstrate the sensitivity of the muon collider to these anomalous couplings, making it an effective tool for investigating CPC and CPV interactions in the Higgs sector.

\section{Event Generation and Selection Procedure}

The $\gamma\gamma \to \gamma\gamma$ scattering process is analyzed within the framework of a high-energy muon collider to impose constraints on anomalous Higgs-gauge boson interactions, with a primary focus on the $H\gamma\gamma$ vertex. Signal events corresponding to non-zero values of the Wilson coefficients $\bar{c}_\gamma$ and $\tilde{c}_\gamma$ are considered, including their interference with the SM contributions. Additionally, a comprehensive set of photon-induced and direct collision backgrounds is simulated to assess their influence on the signal extraction.

The signal includes contributions from effective dim-6 operators, while the main background is the photon-induced $\gamma\gamma\to\gamma\gamma$ (SM background) process, which is the same as the signal process. Another background is considered as the nonphoton-induced process $\mu^+\mu^- \to \gamma\gamma$, which involves direct $\mu^+\mu^-$ collisions. Reducing the background from fake photons to a level well below the irreducible background resulting directly from di-photon production is one of the primary challenges of $H \to \gamma\gamma$ analysis. It is widely known that highly good rejection toward jets is necessary to accomplish this purpose. Therefore, two more background processes where jets may fake two photons are considered; the photon-induced $\gamma\gamma \to jj$ process and the nonphoton-induced $\mu^+\mu^- \to jj$ process. Eventually, a total of four backgrounds are considered to ensure the robustness of the analysis.

Monte Carlo event generation is performed with MadGraph5$\_$aMC@NLO, producing 500k events for each signal and background configuration. Parton showering, fragmentation, and hadronization are simulated using Pythia 8 \cite{Bierlich:2022uzx}, while detector-level responses are modeled with Delphes \cite{Favereau:2014wfb} employing muon collider-specific configuration cards. In addition, ROOT \cite{Brun:1997gqa} is utilized for the subsequent event analysis and statistical evaluations.

\begin{figure}[H]
\centering
\begin{subfigure}{0.5\linewidth}
\includegraphics[width=\linewidth]{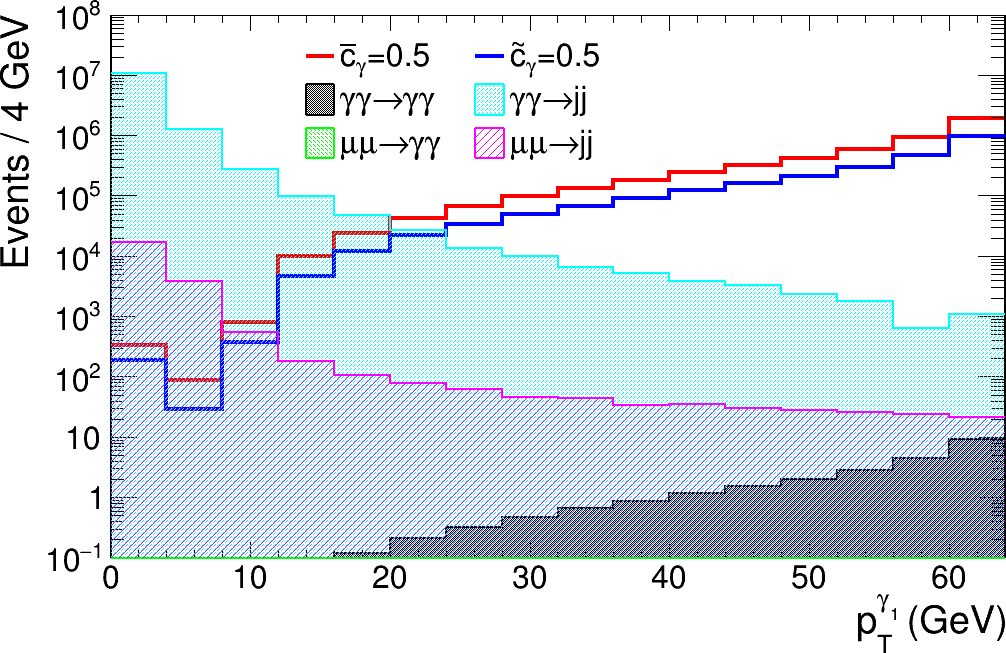}
\caption{}
\label{fig3:a}
\end{subfigure}\hfill
\begin{subfigure}{0.49\linewidth}
\includegraphics[width=\linewidth]{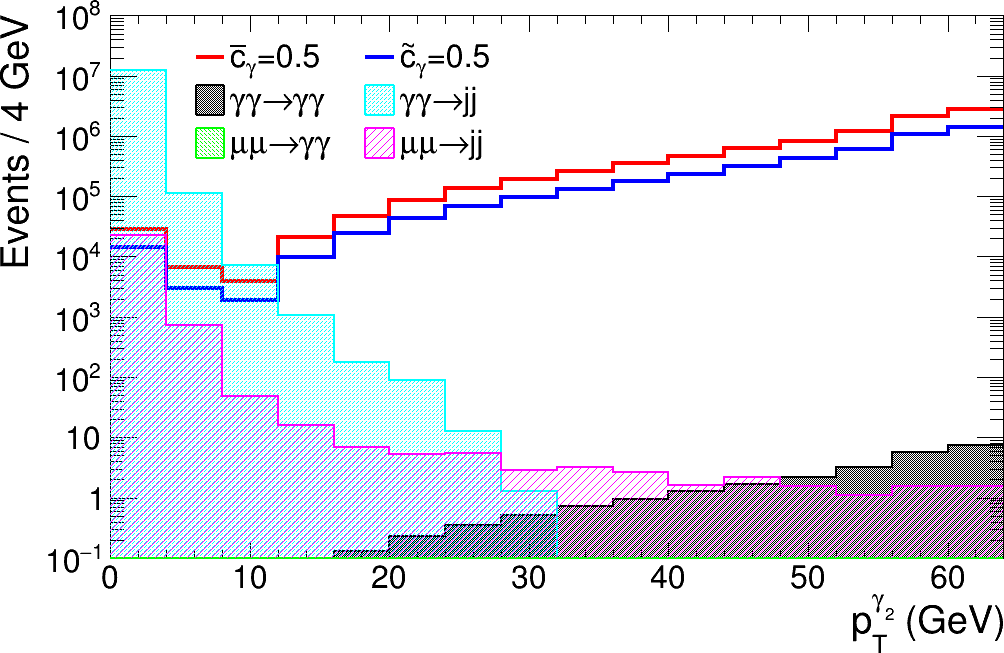}
\caption{}
\label{fig3:b}
\end{subfigure}\hfill
\begin{subfigure}{0.49\linewidth}
\includegraphics[width=\linewidth]{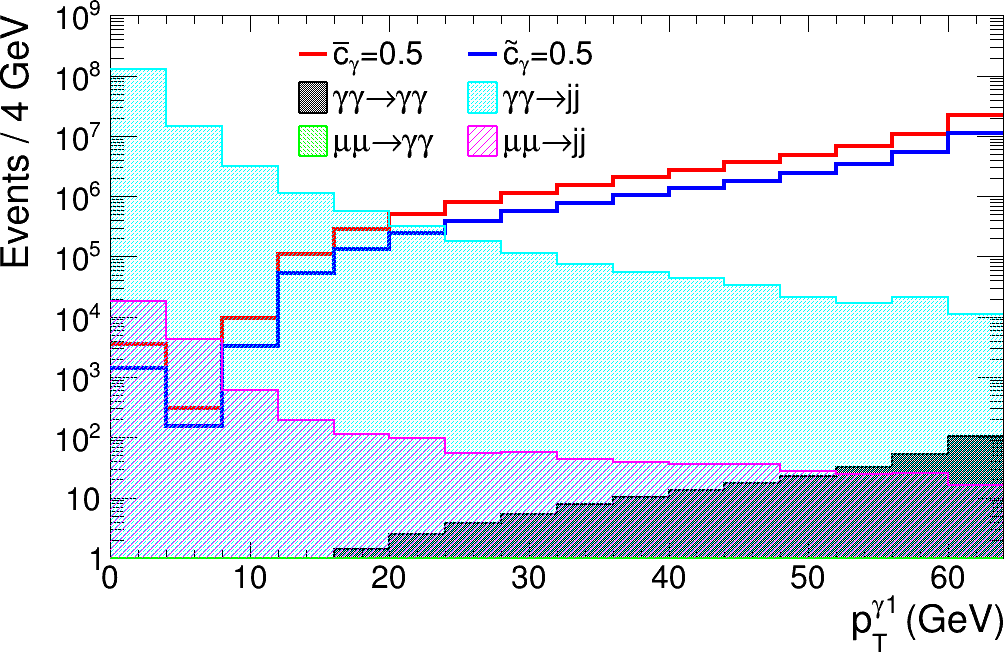}
\caption{}
\label{fig3:c}
\end{subfigure}\hfill
\begin{subfigure}{0.49\linewidth}
\includegraphics[width=\linewidth]{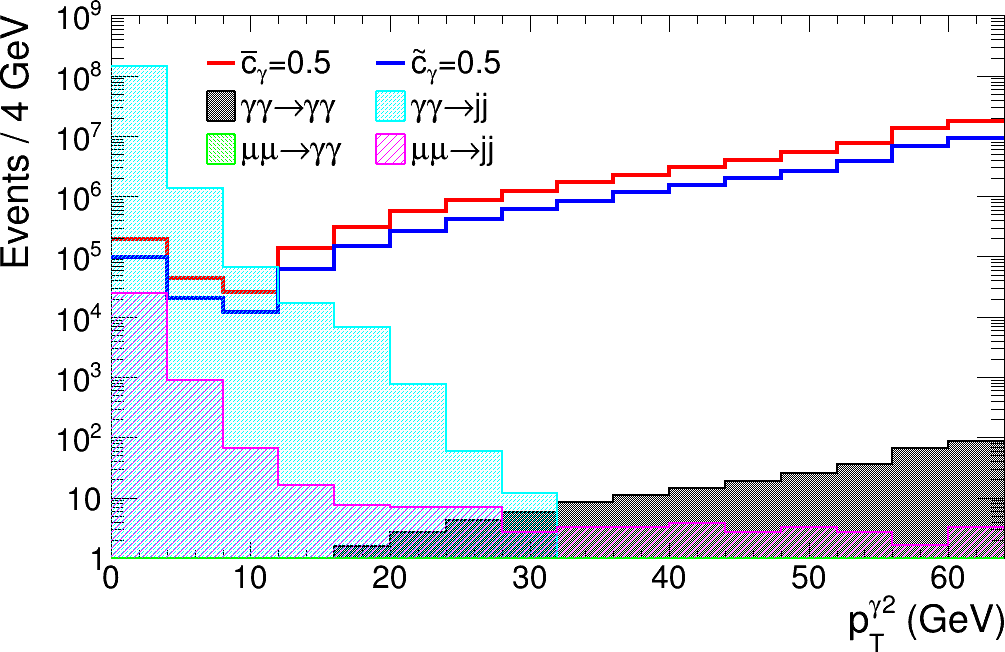}
\caption{}
\label{fig3:d}
\end{subfigure}\hfill
\caption{ \label{fig:3}  The number of events as a function of $p^{\gamma_1}_{T}$ (left) and $p^{\gamma_2}_{T}$ (right) for the process $\gamma\gamma \to \gamma \gamma$ and relevant backgrounds at 10 TeV (top) and 30 TeV (bottom) muon colliders.}
\end{figure}

To optimize the signal-to-background discrimination, a hierarchical preselection and set of kinematic cuts are applied. Initially, for preselection, events containing at least two photons $(N_{\gamma} \geq 2)$ are selected to match the signal topology and the pseudo-rapidity ($\eta^\gamma$) of all photons is set to $|\eta^{\gamma_{1,2}}| < 2.5$. Using with the distributions given in Fig.~\ref{fig:3}, the transverse momentum of the two final state photons $p_T^{\gamma_1}$ and $p_T^{\gamma_2}$ are ordered, with thresholds set at $p_T^{\gamma_1} > 20$ GeV and $p_T^{\gamma_2} > 12$ GeV and are presented with the label Cut-1.

An additional selection criteria is applied based on the minimum distance between the two outgoing photons. Specifically, events are required to satisfy $\Delta R(\gamma_1, \gamma_2) > 3.0$, labeled Cut-2, where $\Delta R$ represents the minimum distance in the pseudo-rapidity - azimuthal angle ($\eta-\phi$) plane. This cut effectively suppresses contributions from background processes where the outgoing photons are closely aligned, particularly those originating from $\gamma\gamma \to jj$ and $\mu^+\mu^- \to jj$, where misidentified jets could mimic photon signatures.

\begin{figure}[H]
\centering
\begin{subfigure}{0.5\linewidth}
\includegraphics[width=\linewidth]{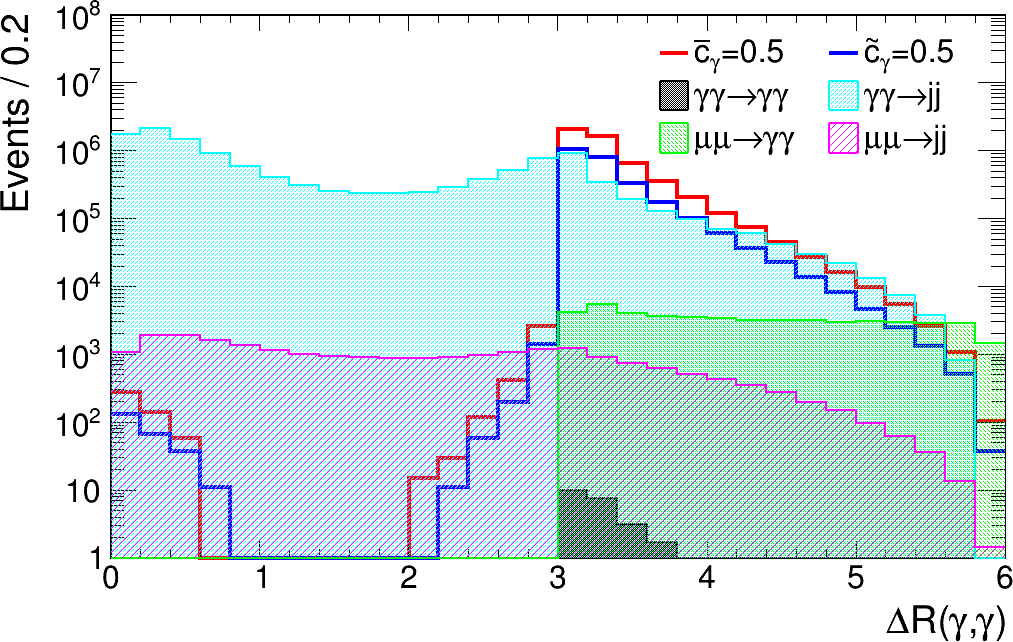}
\caption{}
\label{fig4:a}
\end{subfigure}\hfill
\begin{subfigure}{0.5\linewidth}
\includegraphics[width=\linewidth]{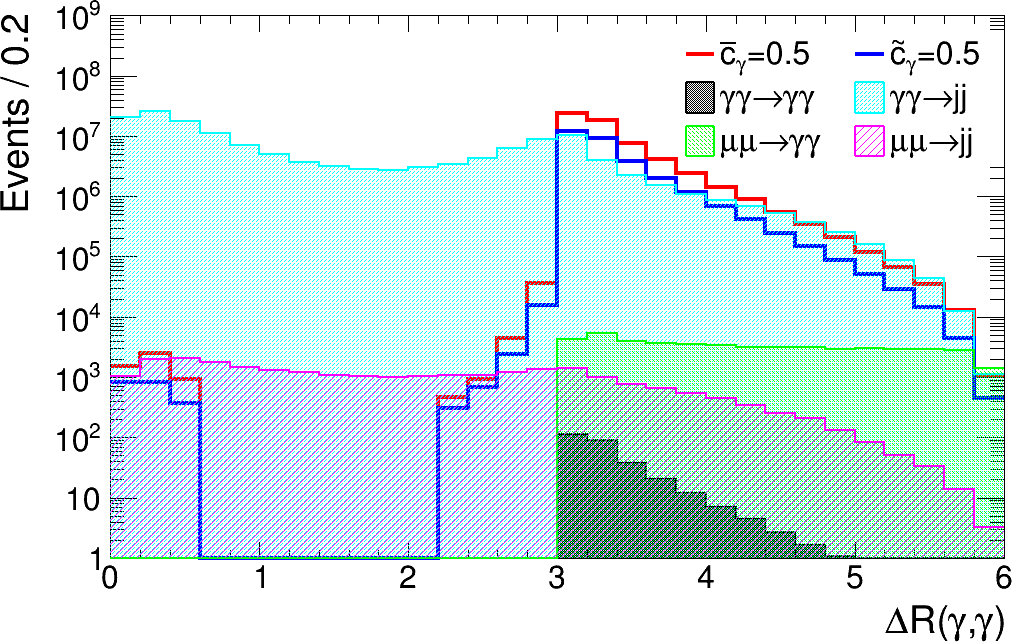}
\caption{}
\label{fig4:b}
\end{subfigure}\hfill
\caption{ \label{fig:4} $\Delta R_{\gamma\gamma}$ distribution for the process $\gamma\gamma \to \gamma \gamma$ and relevant backgrounds at 10 TeV (left) and 30 TeV (right) muon colliders.}
\end{figure}

The distribution of $\Delta R(\gamma_1, \gamma_2)$ for both the signal and relevant background processes is shown in Fig.~\ref{fig:4}. As illustrated, signal events predominantly populate the region with larger $\Delta R$ values compared to backgrounds, allowing for a clear distinction.

\begin{figure}[H]
\centering
\begin{subfigure}{0.5\linewidth}
\includegraphics[width=\linewidth]{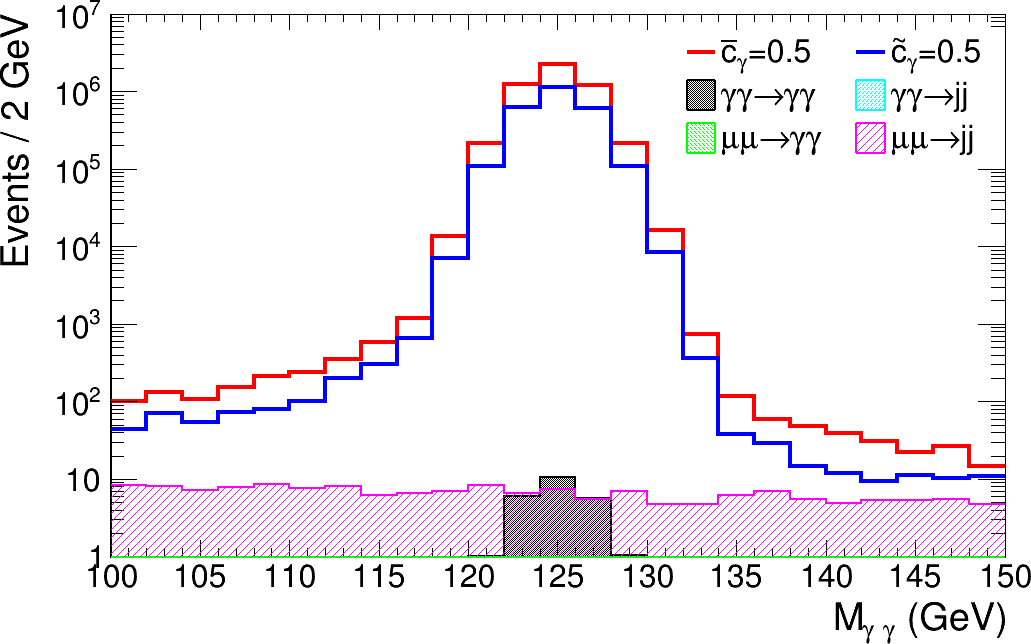}
\caption{}
\label{fig5:a}
\end{subfigure}\hfill
\begin{subfigure}{0.5\linewidth}
\includegraphics[width=\linewidth]{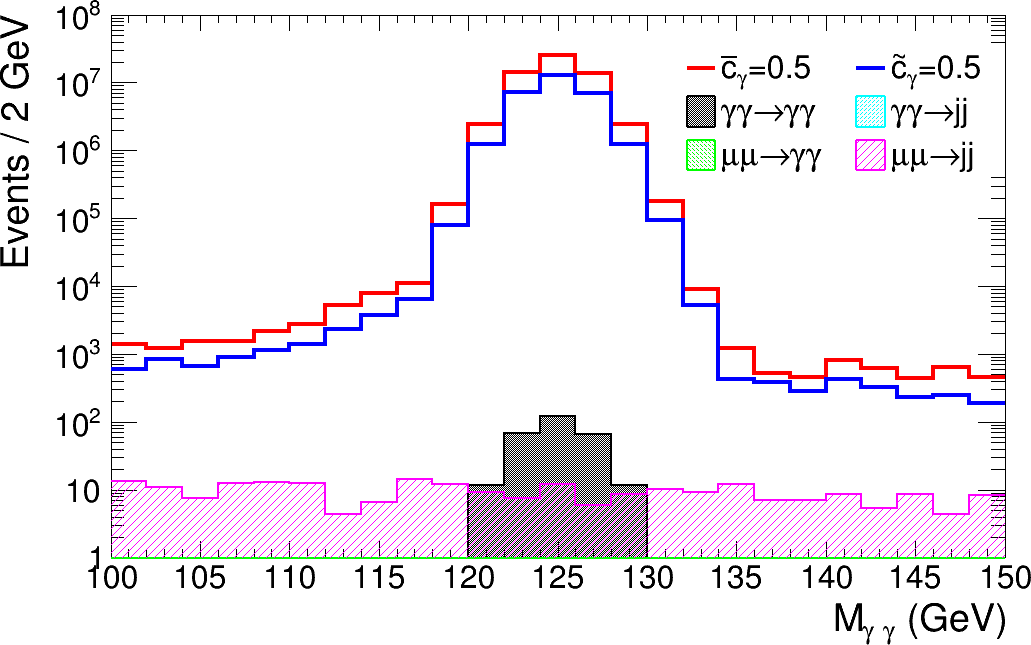}
\caption{}
\label{fig5:b}
\end{subfigure}\hfill
\caption{ \label{fig:5}  Invariant mass $M_{\gamma\gamma}$ distributions of the diphoton system for the process $\gamma\gamma \to \gamma \gamma$ and relevant backgrounds at 10 TeV (left) and 30 TeV (right) muon colliders.}
\end{figure}

Also, an invariant mass selection of $120 < M_{\gamma\gamma} < 130$ GeV is applied to isolate the signal events corresponding to the Higgs boson decay from the background and this selected cut is labeled as Cut-3. This range is centered around the Higgs boson mass $(m_H = 125$ GeV), effectively suppressing non-resonant background contributions. The $M_{\gamma\gamma}$ distributions for signal and background processes are shown in Fig.~\ref{fig:5}, highlighting the ability of this cut to enhance signal purity by focusing on the resonance region.

Since the distributions of the 10 and 30 TeV muon colliders are similar in Figs.~\ref{fig:3}-\ref{fig:5}, the kinematic cuts have been determined common for both colliders. These selection criteria are summarized in Table~\ref{tab1}, providing a structured approach for optimizing signal extraction.

\begin{table}[H]
\centering
\caption{Particle-level selection cuts for the analysis of $\gamma\gamma \to \gamma \gamma$ signal and relevant backgrounds at 10  and 30 TeV muon colliders.}
\label{tab1}
\begin{tabular}{p{5cm}p{5cm}}
\hline
\hline
Kinematic Cuts & Descriptions  \\
\hline
Cut-1   & $p^{\gamma_{1}}_T > 20$ GeV, $p^{\gamma_{2}}_T > 12$ GeV  \\
Cut-2   & $\Delta R(\gamma,\gamma) > 3.0$ \\
Cut-3   & $120<M_{\gamma\gamma}<130$ GeV\\
\hline\hline
\end{tabular}
\end{table}

On the other hand, Table~\ref{tab2} presents the number of events remaining after each cut for the signals ($\bar{c}_\gamma = 0.5, \tilde{c}_\gamma = 0.5$) and the relevant backgrounds $(\gamma\gamma \to \gamma\gamma$, $\mu^+\mu^- \to jj$, $\gamma\gamma \to jj$). The number of events are obtained from the simulated ROOT samples after applying the full analysis chain, including the kinematic selection criteria, parton showering and hadronization with PYTHIA8, and detector simulation with Delphes. The selected events are normalized using the factor

\begin{equation}
\label{eq.7} 
w=\frac{\sigma L_{\rm int}}{N_{\rm gen}},
\end{equation}

{\raggedright where $\sigma$ is the production cross section obtained from MadGraph5$\_$aMC@NLO, $L_{\rm int}$ is the integrated luminosity of the muon collider (e.g. 10 or 90 ab$^{-1}$ depending on the considered stages), and $N_{\rm gen}$ is the total number of generated Monte Carlo events. Consequently, the expected number of events given in Table~\ref{tab2} is calculated as}

\begin{equation}
\label{eq.8} 
N = N_{\rm sel}\times \frac{\sigma L_{\rm int}}{N_{\rm gen}},
\end{equation}

{\raggedright where $N_{\rm sel}$ denotes the number of events surviving all analysis selections. Equivalently, this expression can be written as $N=\sigma L_{\rm int}\epsilon$, with $\epsilon=N_{\rm sel}/N_{\rm gen}$ representing the total selection efficiency.}

As seen from the signal-to-background ratios $S/B_{tot}$ in Table~\ref{tab2}, selected cuts significantly improve the signal-background separation, effectively suppressing background contributions.

\begin{table}[H]
\centering
\caption{Production cross sections ($\sigma$) of the signal and background processes before any event selection, together with the corresponding number of events after each kinematic cut, at 10 TeV and 30 TeV muon colliders.}
\label{tab2}
\begin{tabular}{p{0.4cm}p{1.8cm}p{1.8cm}p{1.4cm}p{1.1cm}p{0.3cm}p{1.4cm}p{1.1cm}p{0.3cm}p{1.4cm}p{1.1cm}p{0.3cm}p{1.4cm}p{1.1cm}}
\hline \hline
& & & \multicolumn{2}{c}{Preselection} && \multicolumn{2}{c}{Cut-1} &&
\multicolumn{2}{c}{Cut-2} && \multicolumn{2}{c}{Cut-3}\\
\hline

& Signal & $\sigma$ [pb] & Events & $S/B_{tot}$ &&
Events & $S/B_{tot}$ &&
Events & $S/B_{tot}$ &&
Events & $S/B_{tot}$\\
\cline{2-14}
\parbox[p]{1cm}{\multirow{6}{*}{\rotatebox[origin=c]{90}{10 TeV Muon Collider}}}
& $\bar{c}_{\gamma}=0.5$
& 0.65744
& 467960 & 0.14501
&& 462997 & 2064.58
&& 462995 & 2613.06
&& 459798 & 203794 \\

& $\tilde{c}_{\gamma}=0.5$
& 0.32888
& 234905 & 0.07279
&& 232460 & 1036.57
&& 232457 & 1311.94
&& 230801 & 102296 \\

\cline{2-2}

& Background & & & & & & & & & & & \\

\cline{2-2}

& $B_{\gamma\gamma \to \gamma\gamma}$
& $3.16\times10^{-6}$ &
2.24907 & &&
2.22512 & &&
2.22508 & &&
2.20949 & \\

& $B_{\gamma\gamma \to jj}$
& 4.51340 &
3223321 & &&
219.222 & &&
174.088 & &&
0 & \\

& $B_{\mu\mu \to jj}$
& 0.00544 &
3718.93 & &&
2.81008 & &&
0.87182 & &&
0.04670 & \\

\hline

& Signal & $\sigma$ [pb] & Events & $S/B_{tot}$ &&
Events & $S/B_{tot}$ &&
Events & $S/B_{tot}$ &&
Events & $S/B_{tot}$\\

\cline{2-14}

\parbox[p]{1cm}{\multirow{6}{*}{\rotatebox[origin=c]{90}{30 TeV Muon Collider}}}
& $\bar{c}_{\gamma}=0.5$
& 0.85212
& 5362842 & 0.13449
&& 5303950 & 1223.78
&& 5303878 & 1535.13
&& 5267285 & 207593 \\

& $\tilde{c}_{\gamma}=0.5$
& 0.41902
& 2693614 & 0.06755
&& 2665957 & 615.116
&& 2665907 & 771.609
&& 2647407 & 104339 \\

\cline{2-2}

& Background & & & & & & & & & & & \\

\cline{2-2}

& $B_{\gamma\gamma \to \gamma\gamma}$
& $4.12\times10^{-6}$ &
25.8449 & &&
25.5539 & &&
25.5534 & &&
25.3731 & \\

& $B_{\gamma\gamma \to jj}$
& 6.20264 &
39870000 & &&
4306.41 & &&
3429.17 & &&
0 & \\

& $B_{\mu\mu \to jj}$
& 0.00061 &
3650.84 & &&
2.10203 & &&
0.27248 & &&
0 & \\

\hline \hline
\end{tabular}
\end{table}

In this study, kinematic cuts are chosen with an optimistic approach, and therefore they suppress some backgrounds very effectively, eliminating the background and reducing the event number to zero. If no background events occur in $n$ independent samples, according to Poisson statistics, the upper limit of the 95\% C.L. is 3 events. This is called the ``rule of three" in statistics \cite{Jovanovic:1997uga}. This statistic has been discussed in many particle physics studies \cite{Wang:2021opv,Zhang:2021dcb}. When zero background events are observed after Cut-3 in the Monte Carlo simulation as shown in Table~\ref{tab2}, the upper limit of the 95\% C.L. according to Poisson statistics is 3 events. Therefore, the maximum survival efficiency ($\epsilon_{max}$) corresponds to the ratio of the upper limit event number to the number of generated Monte Carlo events, $\epsilon_{max}=3/500000=6 \times 10^{-6}$. The maximum background cross-section $\sigma_{max}$ after Cut-3 is $\sigma_{max} = \sigma_{initial} \times \epsilon_{max} = 2.7 \times 10^{-5}$ pb, where $\sigma_{initial}$ is the cross section of the background process $\gamma\gamma \rightarrow jj$ at 10 TeV muon collider before any event selection. When this cross section $\sigma_{max}$ is compared with the signal cross-sections of the coefficients $\bar{c}_\gamma$ and $\tilde{c}_\gamma$ after Cut-3 at 10 TeV muon collider, the signals are approximately 1703 and 855 times larger, respectively. The same comparison can be made for the background processes $\gamma\gamma \rightarrow jj$ and $\mu \mu \rightarrow jj$ at 30 TeV muon collider. The signal cross-sections of the coefficients $\bar{c}_\gamma$ and $\tilde{c}_\gamma$ after Cut-3 at 30 TeV muon collider are approximately 1582 and 795 times larger, respectively, than the cross-section $\sigma_{max}$ of the background process $\gamma\gamma \rightarrow jj$, and approximately 16256 and 8169 times larger, respectively, than the cross-section $\sigma_{max}$ of the background process $\mu \mu \rightarrow jj$. By taking the 95\% C.L. upper limit for the background, we actually compared the worst-case scenario statistically, and yet there is still a large coefficient difference between it and the signal. Thus, even if our Monte Carlo statistics are insufficient, it can be stated with certainty that these backgrounds are unlikely to interfere with the signal.

\section{Projected Sensitivities on Dim-6 Higgs-gauge Boson Couplings}

The sensitivity of the anomalous Higgs-gauge boson couplings in the $\gamma\gamma \to \gamma\gamma$ process is analyzed by employing a $\chi^2$ test to simulated data. To derive constraints at the 95\% Confidence Level (C.L.), the $\chi^2$ distribution is used with  corresponding to one degree of freedom being 3.84. The $\chi^2$ formula is expressed as:

\begin{eqnarray}
\label{eq.9}
\chi^{2} = \sum_{i}^{n_{\text{bins}}} \left( \frac{N_{i}^{\text{TOT}} - N_{i}^{\text{B}}}{N_{i}^{\text{B}} \Delta_{i}} \right)^2,
\end{eqnarray}

{\raggedright where $N_{i}^{\text{TOT}}$ is the total number of events containing contributions from the effective couplings, $N_{i}^{\text{B}}$ represents the number of background events in the $i$-th bin of the photon transverse momentum $p_T^{\gamma\gamma}$ distribution, and $\Delta_{i}$ accounts for the systematic uncertainty $(\delta_{sys})$ and the statistical uncertainty $(\delta_{st}=1/\sqrt{N_i^B})$, defined as $\Delta_{i} = \sqrt{\delta_{sys}^2+\delta_{st}^2}$.}

\begin{figure}[H]
\centering
\begin{subfigure}{0.5\linewidth}
\includegraphics[width=\linewidth]{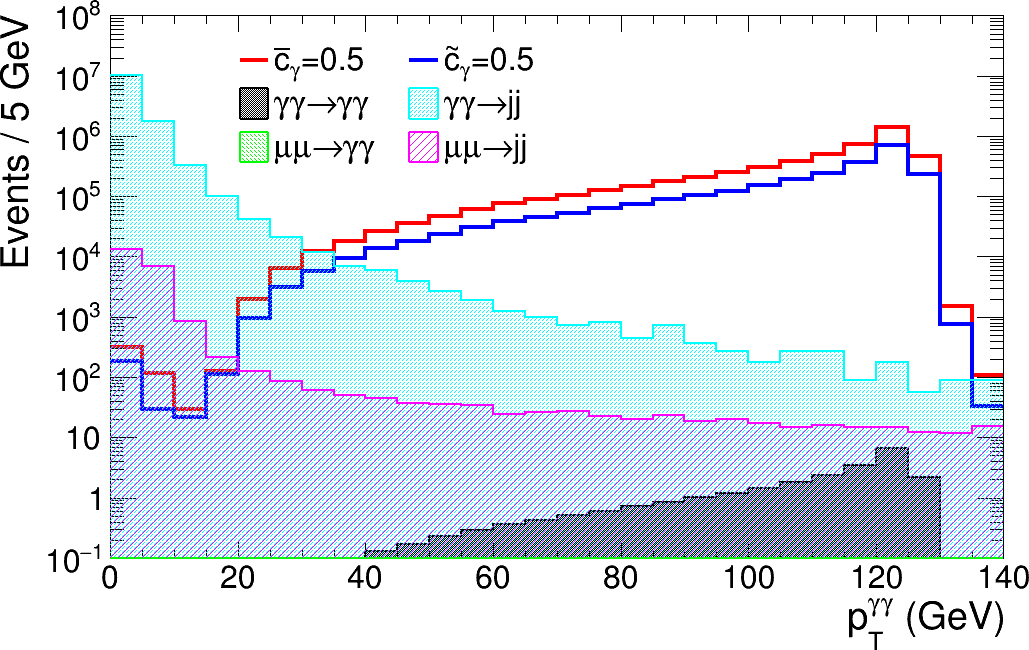}
\caption{}
\label{fig6:a}
\end{subfigure}\hfill
\begin{subfigure}{0.49\linewidth}
\includegraphics[width=\linewidth]{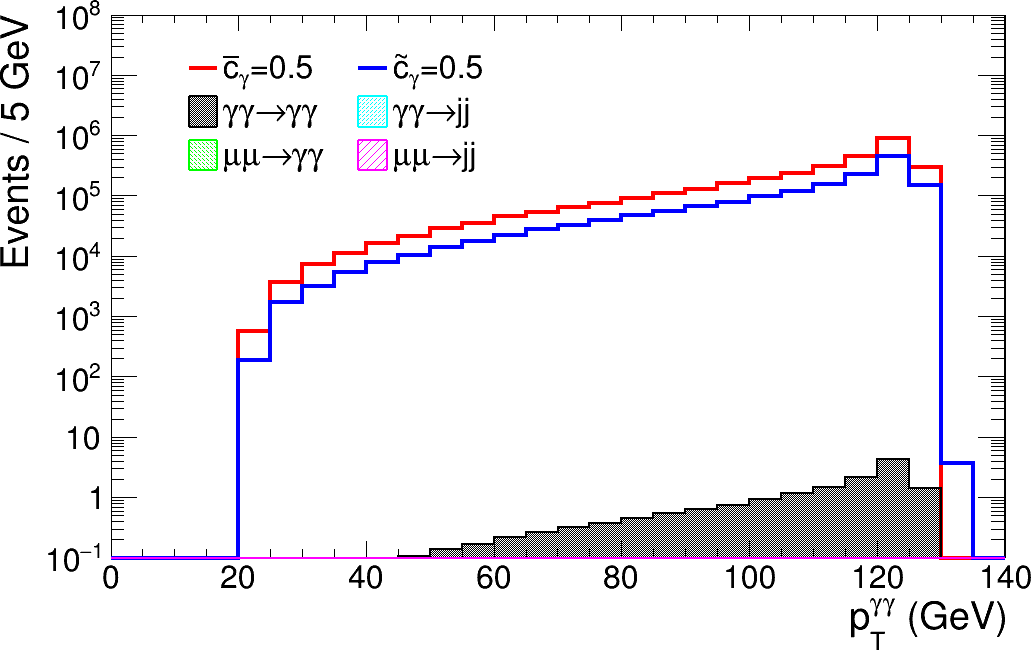}
\caption{}
\label{fig6:b}
\end{subfigure}\hfill
\begin{subfigure}{0.5\linewidth}
\includegraphics[width=\linewidth]{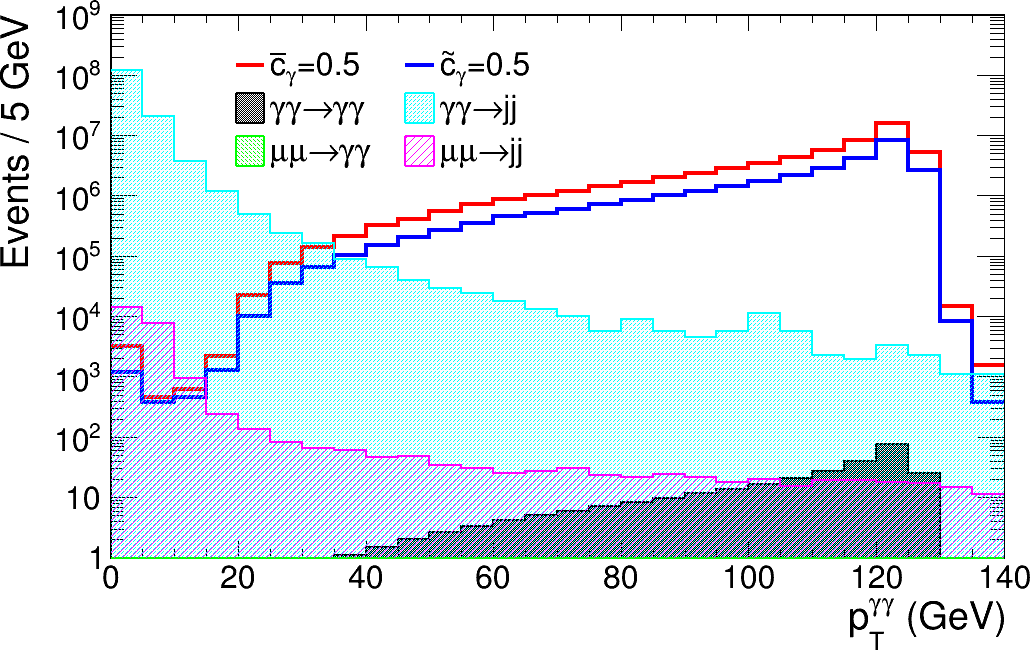}
\caption{}
\label{fig6:c}
\end{subfigure}\hfill
\begin{subfigure}{0.49\linewidth}
\includegraphics[width=\linewidth]{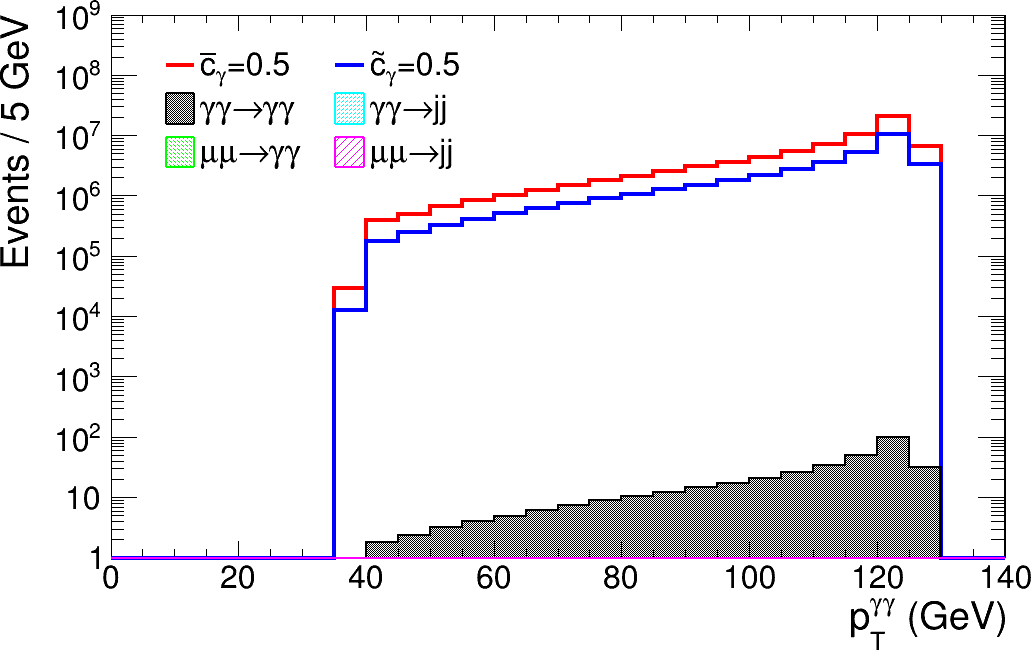}
\caption{}
\label{fig6:d}
\end{subfigure}\hfill
\caption{ \label{fig:6}  Transverse momentum distribution $p^{\gamma\gamma}_{T}$ of the diphoton system (left) before and (right) after selected cuts for the process $\gamma\gamma \to \gamma \gamma$ and relevant backgrounds at 10 TeV (top) and 30 TeV (bottom) muon colliders.}
\end{figure}

While obtaining the sensitivities on $\bar{c}_\gamma$ and $\tilde{c}_\gamma$ coefficients, we have aimed on the transverse momentum of the diphoton system. The transverse momentum distributions of the diphoton system before and after selected cuts are given in Fig.~\ref{fig:6}.

The 95\% C.L. constraints on $\bar{c}_\gamma$ and $\tilde{c}_\gamma$ without and with systematic uncertainties of 5\% and 10\% are determined for a muon collider operating at 10 and 30 TeV CoM energies with 10 and 90 ab$^{-1}$ integrated luminosities, respectively. The obtained sensitivities, related phenomenological and experimental results in ATLAS collaboration are presented in Table~\ref{tab3}. The results highlight the potential of the muon collider to probe these couplings showcasing its sensitivity to both CPC and CPV contributions.

\begin{table}[H]
\centering
\caption{Projected sensitivities at $95\%$ C.L. on the $\bar{c}_\gamma$ and $\tilde{c}_\gamma$ coefficients.}
\label{tab3}
\begin{tabular}{p{3cm}p{1.5cm}p{6cm}p{4.4cm}}
\hline
\hline
Coefficients & & Experimental Results (ATLAS) & Phenomenological Results \\ \hline
\multirow{3}{*}{$\bar{c}_{\gamma}$} & & $[-0.00074;0.00057]$ \cite{Aad:2016hws} & $[-0.0051; 0.0038]$ \cite{Denizli:2019oxc} \\
& & $[-0.00011;0.00011]$ \cite{ATLAS:2019sdf} & $[-0.0089; 0.0066]$ \cite{Denizli:2021uhb} \\
& & & $[-0.0023; 0.0065]$ \cite{Spor:2024sdk} \\ \hline

\multirow{3}{*}{$\tilde{c}_{\gamma}$} & & $[-0.0018;0.0018]$ \cite{Aad:2016hws} & $[-0.0043; 0.0043]$ \cite{Denizli:2019oxc} \\
& & $[-0.00028;0.00043]$ \cite{ATLAS:2019sdf} & $[-0.0077; 0.0077]$ \cite{Denizli:2021uhb} \\
& & & $[-0.0039; 0.0039]$ \cite{Spor:2024sdk} \\ \hline

This study & $\delta_{sys}$ & 10 TeV Muon Collider & 30 TeV Muon Collider \\ \hline
\multirow{3}{*}{$\bar{c}_{\gamma}$} & 0\% & $[-0.000177;0.000215]$ & $[-0.000054;0.000057]$ \\
& 5\% & $[-0.000179;0.000216]$ & $[-0.000060;0.000062]$ \\
& 10\% & $[-0.000183;0.000222]$ & $[-0.000071;0.000074]$ \\ \hline

\multirow{3}{*}{$\tilde{c}_{\gamma}$} & 0\% & $[-0.000921;0.000918]$ & $[-0.000501;0.000486]$ \\
& 5\% & $[-0.000927;0.000924]$ & $[-0.000525;0.000516]$ \\
& 10\% & $[-0.000939;0.000936]$ & $[-0.000576;0.000571]$ \\
\hline \hline
\end{tabular}
\end{table}

\begin{figure}[H]
\centerline{\scalebox{1.5}{\includegraphics{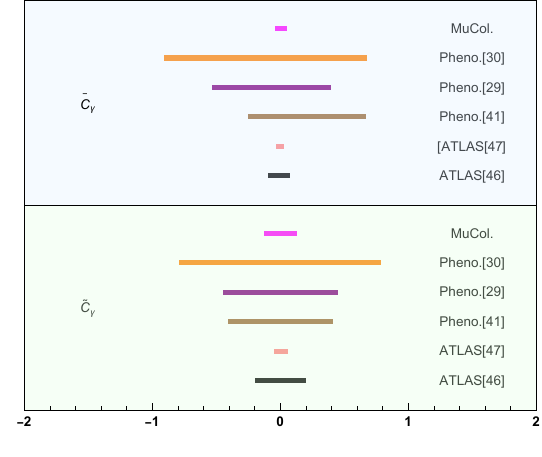}}}
\caption{ \label{fig:7} Comparison of our projected sensitivities on the anomalous $\bar{c}_\gamma$ and $\tilde{c}_\gamma$ coefficients via the process $\gamma\gamma \to \gamma \gamma$ at future muon collider with other sensitivities. Here, the x-axis illustrates ``Couplings$\times 10^{-2}$".}
\end{figure}

Fig.~\ref{fig:7} presents a comparison between the obtained sensitivities for the future muon collider and the experimental results reported by ATLAS and some phenomenological studies, illustrated through bar charts. The results in Fig.~\ref{fig:7} highlight the capability of future muon colliders to explore new physics scenarios with exceptional precision, surpassing current experimental limits.

\begin{figure}[H]
\centerline{\scalebox{0.5}{\includegraphics{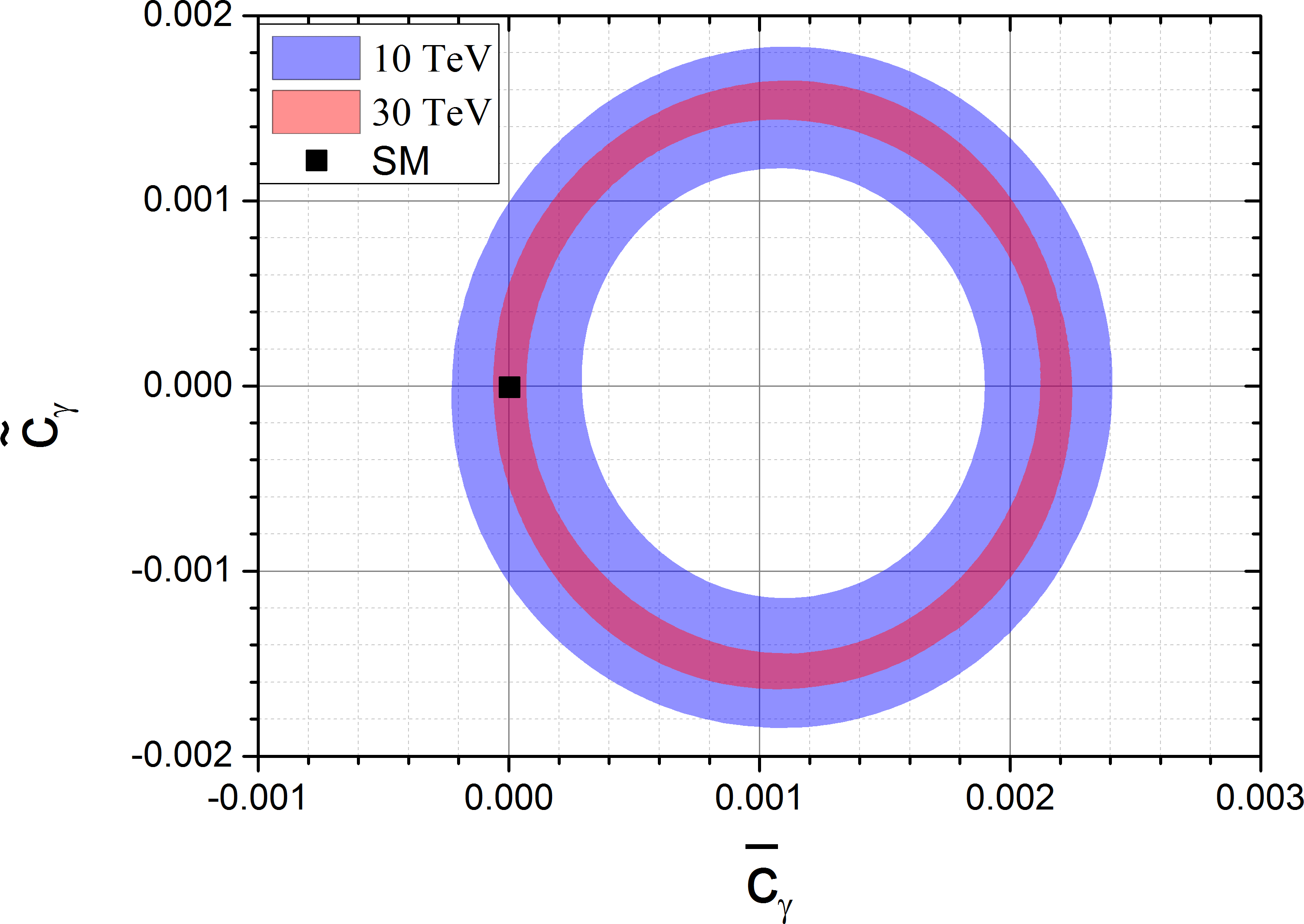}}}
\caption{ \label{fig:8} Two-dimensional 95\% C.L. intervals in plane for $\bar{c}_\gamma$ and $\tilde{c}_\gamma$ at muon colliders with $\sqrt{s}=10$ TeV (${\cal L}_{int}=10$ ab$^{-1}$) and $\sqrt{s}=30$ TeV (${\cal L}_{int}=90$ ab$^{-1}$). The black square represents the SM expectation.}
\end{figure}

In addition to single-parameter analyses, two-parameter scenarios are also examined where $\bar{c}_\gamma$ and $\tilde{c}_\gamma$ coefficients are variable and the critical values of $\chi^2$ corresponding to two degrees of freedom are equal to 5.99 by applying the $\chi^2$ test. Two-dimensional 95\% C.L. contours are generated for the $\bar{c}_\gamma$-$\tilde{c}_\gamma$ parameter space, providing a clear visualization of how CPC and CPV contributions to the $H\gamma\gamma$ vertex can be constrained together. When planning and designing a muon collider with a center-of-mass energy of 10 TeV or more, a 30 TeV stage appears to be an attractive option. These results, presented in Fig.~\ref{fig:8}, demonstrate the interaction of the two coefficients, the evolution between the two different muon collider stages and the future muon collider's ability to set bounds on their combined effects.

\section{Conclusions}

The process $\gamma\gamma \to \gamma\gamma$ is examined to establish constraints on the Wilson coefficients $\bar{c}_\gamma$ and $\tilde{c}_\gamma$ at the future muon collider with CoM energies of 10 and 30 TeV. The analysis utilizes the Weizsäcker-Williams approximation to model incoming photons. Signal events with non-zero couplings, including their interference with the SM and background events, are generated using MadGraph, where the SMEFT effective Lagrangian is implemented through FeynRules and the UFO framework. 

Subsequently, the events are processed with PYTHIA 8 for parton showering and hadronization, followed by Delphes to account for detector response. The analysis focuses on a signature of the diphoton final state. Kinematic distributions, such as the transverse momentum, the invariant mass and the minimum distance, are analyzed to optimize the extraction of signal events and suppress backgrounds using cut-based approach on the final-state photons.

Finally, a $\chi^2$-based analysis is performed to extract limits on the anomalous couplings. Our best obtained sensitivities on $\bar{c}_\gamma$ and $\tilde{c}_\gamma$ at the 30 TeV muon collider are $[-0.00005;0.00005]$ and $[-0.00050;0.00048]$, respectively. The two-parameter analysis reveals that the muon collider stage with a CoM energy of 30 TeV achieves stringent bounds for the $\bar{c}_\gamma$ and $\tilde{c}_\gamma$ coefficients than the stage with a CoM energy of 10 TeV. For the 30 TeV muon collider scenario, the obtained constraints on $\bar{c}_\gamma$ and $\tilde{c}_\gamma$ have higher precision than many experimental and phenomenological results in the literature.

\section{Data Availability Statement}

This manuscript has no associated data or the data will not be deposited. [Authors’ comment: Data will be made available upon reasonable request.]

\end{document}